\documentclass{elsart}

\usepackage{amsmath}
\usepackage{amssymb}
\usepackage{graphicx}% include figure files
\usepackage{bbm}% blackboard bold math

\def\R{{\mathbbm{R}}}

\begin{document}

\begin{frontmatter}



\title{When topology triggers a phase transition}


\author{Michael Kastner}

\address{Physikalisches Institut, Lehrstuhl f\"ur Theoretische Physik I,\\ \small Universit\"at Bayreuth, 95440 Bayreuth, Germany}

\ead{Michael.Kastner@uni-bayreuth.de}

\begin{abstract}
Two mathematical mechanisms, responsible for the generation of a thermodynamic singularity, are individuated. For a class of short-range, confining potentials, a topology change in some family of configuration space submanifolds is the {\em only}\/ possible such mechanism. Two examples of systems in which the phase transition is {\em not}\/ accompanied by a such topology change are discussed. The first one is a model with long-range interactions, namely the mean-field $\varphi^4$-model, the second example is a one-dimensional system with a non-confining potential energy function. For both these systems, the thermodynamic singularity is generated by a maximization over one variable (or one discrete index) of a smooth function, although the context in which the maximization occurs is very different.
\end{abstract}

\end{frontmatter}

\section{Introduction}
Phase transitions, like the boiling and evaporating of water at a certain temperature and pressure, are common phenomena both in everyday life and in almost any branch of physics. Loosely speaking, a phase transition brings about a sudden change of the macroscopic properties of a system while smoothly varying a parameter (the temperature or the pressure in the above example). The mathematical description of phase transitions is conventionally based either on Gibbs measures on phase space or on (grand)canonical thermodynamic functions, relating their loss of analyticity (or, in other words, the appearance of a singularity) to the occurrence of a phase transition.

Resorting to the established formalism of equilibrium statistical mechanics, a singularity in, say, the canonical free energy density appears or not, depending on the system under investigation, somewhat as from a black box: the Hamiltonian of the system is put into the definition of a partition function, resulting in a singularity or not. But what is the origin of such a singularity? This fundamental question may be considered as settled, but it is elusive at the same time, depending on the {\em level}\/ at which one tries to answer it. Following a suggestion by Hendrik A.\ Kramers \cite{Kramers:37} and the subsequent more elaborate confirmation by Chen N.\ Yang and Tsung D.\ Lee \cite{LeeYang1}, the thermodynamic limit of the number of degrees going to infinity can be identified as the mechanism generating a thermodynamic singularity.

In the present article, the origin of a thermodynamic singularity is discussed on a different level, leading to a more differentiated result: two mechanisms, very unlike in their mathematical structure, are discussed with particular emphasis on the classes of systems for which they can occur. The first mechanism, set forth in Sec.\ \ref{sec:topology}, is related to topology changes within a certain family of submanifolds in configuration space, whereas the second mechanism, discussed in Sec.\ \ref{sec:maximization}, is based on a maximization over one variable of a smooth function. The different mechanisms allow to classify systems accordingly, and the classification is reflected to some extend in the universality classes.

\section{Singularities from topology changes}
\label{sec:topology}
The use of concepts from topology to describe a physical phenomenon is particularly appealing due to the fact that topology constitutes a very reductional description: considering only the topology of, say, a surface, a significant amount of information is disregarded. If one then succeeds to capture the essentials of the phenomenon of interest with the remaining information only, the desirable goal of an efficient description has been achieved. It was in the late 1990s when studies of the geometric structure of Hamiltonian dynamics led to a conjectured relation between certain topological quantities and phase transitions \cite{CaCaClePe:97,CaPeCo:00}. Although later this hypothesis was found not to be valid in its generality, it has been proved rigorously for a certain class of systems \cite{FraPe:04,FraPeSpi}.

Consider a system of $N$ classical degrees of freedom, characterized by the Hamiltonian function
\begin{equation}
H=\frac{1}{2}\sum_{i=1}^N p_i^2 + V(q)
\end{equation}
consisting of a standard kinetic energy term quadratic in the momenta $p_i$, and a potential energy $V(q)$, depending on the position coordinates $q=(q_1,\dotsc,q_N)\in\Gamma$ from some {\em continuous}\/ configuration space $\Gamma$ (typically a subset of $\R^N$). Then we define a family of submanifolds $\left\{M_v\right\}_{v\in\R}$, where
\begin{equation}
M_v=\left\{q\in\Gamma\,\big|\,V(q)\leqslant Nv\right\}
\end{equation}
is the set of all points from configuration space $\Gamma$ with potential energy equal to or below the threshold value $Nv$. We speak of a {\em topology change}\/ at some value $v_t$ whenever $M_{v-\epsilon}$ and $M_{v+\epsilon}$ are not homeomorphic for arbitrarily small positive $\epsilon$.

A connection between the topology of the $M_v$ and phase transitions was established by noting that, for some toy models studied, a phase transition at a critical energy $v_c$ is accompanied by a topology change within the family of manifolds $\left\{M_v\right\}$ at $v=v_c$. For a class of systems with smooth, bounded below and confining potential $V$ of finite range, Franzosi, Pettini, and Spinelli \cite{FraPe:04,FraPeSpi} were able to prove a theorem stating, loosely speaking, that a topology change within the $\left\{M_v\right\}$ at $v=v_c$ is {\em necessary}\/ for a phase transition to take place at a critical potential energy $v_c$. From these results we can identify a topology change within the configuration space submanifolds $M_v$ as one possible mechanism to generate a thermodynamic singularity. The above cited theorem then asserts that, for the class of systems covered by its assumptions, a topology change is the {\em only}\/ such mechanism that can occur.

The observed connection between topology and phase transitions led to the conjectures that such a connection might exist for general systems \cite{CaCaClePe:97,CaPeCo:00} and that the phase transition might be characterized from topological information \cite{CaPeCo:03}. Recent results \cite{BaroniPhD,GaSchiSca:04,Kastner:04,AnRuZa:05} reveal that this hope was too optimistic. This leads us to the study of further mechanisms generating a thermodynamic singularity.

\section{Singularities from maximization}
\label{sec:maximization}
In this section, two model systems are discussed for which the phase transition occurring was found not to be related to any of the topology changes. For both systems, a maximization is singled out as the mechanism relevant for the generation of a singularity, although in each case in a very different context.

{\bf Example 1: mean-field $\boldsymbol{\varphi^4}$-model ---}
This model is characterized by the potential energy function
\begin{equation}
V(q)=-\frac{J}{2N}\left(\sum_{i=1}^N q_i\right)^2 + \sum_{i=1}^N \left(-\frac{1}{2}q_i^2 + \frac{1}{4}q_i^4\right),\qquad q\in\R^N,
\end{equation}
where the first term describes mean field-type interactions, coupling each degree of freedom $q_i$ to each other at equal strength $J$. The second term is an on-site potential, having the shape of a double well. Mean field-type interactions are an extreme case of long-range interactions, so the system clearly does not fulfill the requirements of the theorem in Ref.\ \cite{FraPe:04,FraPeSpi}.

For this model both, the thermodynamic behaviour and the configuration space topology have been analyzed. The critical energy $v_c(J)$ at which the system undergoes a phase transition diverges with increasing coupling strength $J$ \cite{GaSchiSca:04,HaKa:05}. In contrast, the topology changes within the family $\left\{M_v\right\}$, although very many, are found to take place at non-positive energies $v\leqslant0$ for arbitrary values of $J$ \cite{BaroniPhD,GaSchiSca:04}. This result clearly excludes the coincidence of the energy of the phase transition with that of any of the topology changes, disproving a general connection between these two quantities for long-range systems (and thus the hypothesis on a general relation between topology and phase transitions put forward in Ref.\ \cite{CaCaClePe:97,CaPeCo:00}).

To learn more about the origin of the thermodynamic singularity in the absence of a topology change, the mean-field $\varphi^4$-model was analyzed within the microcanonical ensemble by means of a large deviation technique in Ref.\ \cite{HaKa:05}. In particular, the microcanonical entropy $s$ was computed in the thermodynamic limit $N\to\infty$, first as a function of potential energy $v$ and magnetization $m$, and second as a function of $v$ only. The function $s(v,m)$ is found to be smooth on its entire domain, and no singularity occurs. It is not until the maximization is performed when computing $s(v)=\sup_m s(v,m)$ that a singularity (in the sense of a discontinuity in some derivative) shows up in the entropy. This maximization is identified as a second mechanism, along with topology changes as discussed in Sec.\ \ref{sec:topology}, responsible for the generation of thermodynamic singularities.

Some comments on this mechanism: A smooth function of $n$ variables resulting in a {\em non-smooth}\/ function of $n-1$ variables upon maximization over one variable has to be non-concave. Since the entropy of well-behaved (stable and tempered) short-range systems is a concave function, the above described mechanism of singularity generation is restricted to systems with long-range interactions. Furthermore, since a Taylor expansion of a smooth $s(v,m)$ around the phase transition point is possible, so-called classical (or mean-field) critical exponents are obtained generically in case of a continuous phase transition (see appendix of Ref.~\cite{HaKa:05} for details).

{\bf Example 2: Burkhardt model ---}
This one-dimensional model, introduced in Ref.\ \cite{Burkhardt:81} to model domain wall fluctuations, is characterized by a potential energy function
\begin{equation}
V(q)=\sum_{i=1}^N \left[\left|q_{i+1}-q_i\right|+U(q_i)\right],\qquad q\in\left(\R^+\right)^N.
\end{equation}
The interactions---in contrast to our first example---are of short range, being restricted to nearest neighbours on the lattice. The on-site potential $U$ is finite, has a single minimum somewhere on its domain (the positive half-line), and it approaches a finite value in the limit of large arguments, $\lim_{x\to\infty}U(x)<\infty$. (Think of something like a Lennard-Jones potential.) Such a potential $V$ is not confining, so this time another one of the requirements of the theorem in Ref.\ \cite{FraPe:04,FraPeSpi} is not met.

Analyzing the configuration space topology of this model \cite{Kastner:04} and comparing the result to the thermodynamic behaviour, the energy of the phase transition, as in the mean-field $\varphi^4$-model, is found to differ from the energy at which the only topology change in the manifolds $M_v$ occurs \cite{AnRuZa:05}. As argued above, due to the concavity of the entropy function, the thermodynamic singularity in a short-range system cannot stem from a maximization over one variable of a smooth entropy function. But, again, the thermodynamic singularity can be traced back to a maximization over smooth functions, however in a completely different context.

Making use of the transfer matrix technique \cite{KraWan:41}, the canonical free energy density $f$ as a function of the inverse temperature $\beta$ of the one-dimensional model can be written as
\begin{equation}
-\beta f(\beta)=\sup_i \ln \lambda_i(\beta).
\end{equation}
By $\lambda_i$ we denote the eigenvalues of the so-called transfer matrix, which can be deduced from the potential energy. Under suitable conditions \cite{Kastner_inprep}, the $\lambda_i$ are smooth functions of $\beta$, and a phase transition can occur when the largest and the second-largest eigenvalue cross. As for the mean-field $\varphi^4$-model, although in a different context, the thermodynamic singularity is generated by a maximization over smooth functions, likewise leading to the generic occurrence of classical critical exponents. 

\section{Conclusions}
Trying to understand in more detail the origin of a phase transition, we have identified two mechanisms which can generate a thermodynamic singularity: a topology change within the family $\left\{M_v\right\}$ of configuration space submanifolds, and a maximization over one variable (or one discrete index) of a smooth function. For a class of short-range, confining potentials, a topology change is the {\em only}\/ possible mechanism to generate a singularity.

Two examples of systems in which the phase transition is {\em not}\/ accompanied by a topology change are discussed. The first one is the mean-field $\varphi^4$-model, where a singularity is generated by the maximization $s(v)=\sup_m s(v,m)$ from a smooth entropy $s(v,m)$, and this mechanism is restricted to long-range systems for which non-concave entropy functions can occur. The second example is the Burkhardt model, a one-dimensional system with non-confining potential. Here, the free energy density can be expressed as a maximization over the eigenvalues of a transfer matrix. Given a maximization over one variable of a {\em smooth}\/ function, the critical exponents will generically have classical (=mean-field) values, so in this sense the singularity generating mechanism is mirrored to some extend in the universality class of the system.

\section*{Acknowledgments}
Financial support by the Deutsche Forschungsgemeinschaft (grant KA2272/2) is acknowledged.

\end{document}